\documentclass[a4paper,11pt]{article}
\usepackage{jcappub} 
\usepackage{lineno}

\usepackage{graphicx}
\usepackage{dcolumn}
\usepackage{bm}
\usepackage{amssymb}
\usepackage{latexsym}
\usepackage{booktabs}
\usepackage{amsmath}
\usepackage{multirow}
\usepackage{url}
\usepackage{hyperref}
\usepackage{footnote}
\usepackage{float}
\usepackage{threeparttable}
\usepackage{color}
\usepackage{array}
\usepackage{enumerate}
\usepackage{adjustbox}
\usepackage{BOONDOX-cal}

\usepackage{subcaption}
\usepackage{makecell}
\usepackage{diagbox}
\usepackage{epstopdf}
\usepackage{epsfig}
\usepackage{longtable}
\usepackage{supertabular}
\usepackage{algorithm}
\usepackage{pifont}
\usepackage{algorithmic}
\usepackage{changepage}
\usepackage{setspace}
\usepackage{booktabs}    
\usepackage{multirow}   
\usepackage{adjustbox}   
\usepackage{caption}

\newcommand{\blue}[1]{\textcolor{blue}{#1}}

\begin{document}	

	\title{Prospective sensitivity of CTAO on detection of evaporating primordial black holes}
        \author[a]{Chen Yang,}
	\author[a]{Jun-Da Pan}
	\author[a,b,c,\href{note}{\ast}]{and Xin Zhang\note[$\blue \ast$]{Corresponding author.}}

	\affiliation[a]{Liaoning Key Laboratory of Cosmology and Astrophysics, College of Sciences, Northeastern University, Shenyang 110819, China}
	\affiliation[b]{National Frontiers Science Center for Industrial Intelligence and Systems Optimization, Northeastern University, Shenyang 110819, China}
	\affiliation[c]{MOE Key Laboratory of Data Analytics and Optimization for Smart Industry, Northeastern University, Shenyang 110819, China}
        \emailAdd{chenyang@stumail.neu.edu.cn}
	\emailAdd{panjunda@stumail.neu.edu.cn}
	\emailAdd{zhangxin@mail.neu.edu.cn}

	\abstract{As the lifetime of a black hole decreases, the energy of the Hawking radiation it emits increases, ultimately culminating in its disappearance through a powerful burst of gamma rays. For primordial black holes (PBHs) with an initial mass of $\sim 5\times10^{14}$ g, their lifespans are expected to end in the present epoch. Detecting such PBH bursts would provide compelling evidence of their existence. The Cherenkov Telescope Array Observatory (CTAO) has the potential to observe these bursts at the high-energy end of the gamma-ray spectrum. To investigate this possibility, we conduct a study to evaluate the sensitivity of CTAO to the local burst rate density of PBHs. Our results suggest that during a 5-year observational campaign, CTAO could exclude a local burst rate density exceeding $\sim 36\ \mathrm{pc}^{-3}\ \mathrm{yr}^{-1}$, which represents an improvement of one order of magnitude over the upper limit set by the Large High Altitude Air Shower Observatory (LHAASO). In addition, we propose an observation strategy optimized for detecting PBH bursts.}
	\maketitle

\section{Introduction}

Primordial black holes (PBHs) are theorized to have formed through the direct gravitational collapse of density fluctuations under extreme conditions in the early universe \cite{Hawking:1971ei}. Unlike black holes formed from stellar collapse, PBHs could have emerged in the very early stages of cosmic evolution, with masses ranging from micrograms to thousands of times the mass of the Sun. Due to their formation mechanism being closely related to the physical conditions of the early universe, PBHs are considered a unique probe for studying cosmology, gravitational theory, and particle physics \cite{Carr:2020gox}. In particular, smaller PBHs are expected to gradually evaporate through the Hawking radiation process, potentially releasing high-energy gamma rays, providing a possible observational window for detecting these enigmatic objects. Understanding the properties and abundance of PBHs not only helps reveal the physical processes of the early universe but may also offer critical insights into the nature of dark matter \cite{Carr:2021bzv}. Therefore, exploring the observational signatures of PBHs and their distribution in the universe has become one of the key topics in modern astrophysics. However, it should be noted that PBHs with masses $\sim 5 \times 10^{14}$ g, while potentially detectable through their evaporation signatures, cannot constitute the entirety of dark matter due to existing observational constraints that limit their contribution to a small fraction of the total dark matter density \cite{Green2024,Carr2024}. Nevertheless, even a subdominant PBHs population could provide valuable insights into early universe physics. For PBHs in the mass window $\sim 10^{14}$--$10^{15}$\,g, multiple observational approaches have been proposed to place constraints. Microlensing surveys primarily probe higher-mass PBHs~\cite{2023ApJ...954..172E}, while cosmological observations provide bounds across broader mass ranges~\cite{2024JCAP...07..003A,Villanueva_Domingo_2021}. Other evaporation-based searches target similar mass ranges~\cite{luque_2025_v611c-abv20}, each employing different assumptions and detection strategies. The PBHbounds database~\cite{PBHbounds} and relevant reviews~\cite{Green2024,Carr2024} offer a comprehensive overview of current observational limits across a wide range of PBH masses and methods.

Hawking radiation, a theoretical prediction by Stephen Hawking, proposes that black holes can emit radiation due to quantum effects near their event horizons, resulting in gradual mass loss over time \cite{Hawking:1974rv}. This radiation typically consists of a variety of fundamental particles \cite{MacGibbon:1990zk, MacGibbon:1991tj}. For PBHs, this process is particularly significant for those with masses below $5\times10^{14}$ g, as they would have fully evaporated by the present epoch \cite{Ukwatta:2015iba}. During the final stages of evaporation, PBHs are expected to emit a burst of high-energy gamma rays, peaking at energies detectable by modern gamma-ray observatories. This explosive emission serves as a unique observational signature, potentially enabling the identification of PBHs. Detecting such gamma-ray signals could not only confirm the existence of PBHs but also provide insights into their mass distribution and abundance in the universe. The Cherenkov Telescope Array Observatory (CTAO), with its superior sensitivity to very-high-energy gamma rays, is poised to play a crucial role in this search, potentially uncovering evidence of PBH evaporation and shedding light on their cosmological implications \cite{CTAConsortium:2017dvg,Knodlseder:2020onx,CTAObservatory:2022mvt}.

The search for PBH burst phenomena characterized by the sudden release of high-energy cosmic rays during their final evaporation stage holds significant scientific importance with the potential to impact multiple research fields. Detecting such bursts would not only provide direct evidence of PBHs but also enable the reconstruction of their spatial distribution through analyses of burst rate density. In contrast, the absence of observed signals would impose stringent upper limits on the local PBH burst rate density. Regardless of the outcome, these searches could constrain the power spectrum of primordial density fluctuations at scales much smaller than those probed by the cosmic microwave background (CMB) \cite{Wang:2019kaf}, offering insights into various dynamical scenarios of the inflationary universe \cite{Carr:2021bzv,Carr:2005zd,Carr:2009jm}. Additionally, the detection of PBH bursts could provide valuable information on quantum gravitational effects and high-energy particle physics beyond the reach of the Large Hadron Collider (LHC) \cite{Evans:2008zzb}, potentially opening new avenues for physics beyond the Standard Model.

In the search for PBH bursts on parsec scales, several existing experiments including the Whipple 10 m telescope (Whipple) \cite{Linton:2006yu}, CYGNUS air-show array (CYGNUS) \cite{Alexandreas:1993zx}, High Energy Stereoscopic System (H.E.S.S.) \cite{HESS:2023zzd}, Tibet Air Shower Array \cite{amenomori1995search},  Milagro high energy observatory (Milagro) \cite{Abdo:2014apa}, Very Energetic Radiation Imaging Telescope Array System (VERITAS) \cite{Archambault:2017asc}, Fermi Large Area Telescope (Fermi-LAT) \cite{Fermi-LAT:2018pfs}, High Altitude Water Cherenkov Observatory (HAWC) \cite{HAWC:2019wla}, Southern Wide field of view Gamma-ray Observatory (SWGO) \cite{Lopez-Coto:2022tcg} and Large High Altitude Air Shower Observatory (LHAASO) \cite{Yang:2024vij} have already set upper limits on the number density or local burst rate of PBHs. 
Previous review work systematically summarizes these IACT constraints, from the early SGARFACE trigger on Whipple to recent 5000-h H.E.S.S. analyses, and discusses prospects for CTAO and SWGO~\cite{Mukherjee_2018}. In Ref.~\cite{Cassanyes:2015wpr}, the MAGIC performance was extrapolated to CTAO, and it was projected that the latter could reach $\mathcal{O}(10^{2})\,\mathrm{pc^{-3}\,yr^{-1}}$ sensitivity thanks to its $\sim$790-times larger effective search volume.
However, these instruments face certain limitations in sensitivity and detection capabilities. In comparison, the CTAO offers significant advantages. CTAO not only covers a wider energy range (20 GeV to 300 TeV) but also exhibits a substantial improvement in sensitivity over existing instruments, with a projected sensitivity increase of approximately one order of magnitude around 1 TeV \cite{CTAConsortium:2017dvg}. Furthermore, CTAO lower energy threshold enables the detection of objects at greater distances, while its higher energy coverage extends the observational reach of ultra-high-energy (UHE) gamma-ray astronomy. Consequently, CTAO is expected to become an ideal platform for detecting PBH bursts, with the potential to either achieve the first detection of PBH bursts or provide the most stringent upper limits on the local burst rate density. In summary, the capabilities of CTAO in the detection of PBHs present new opportunities for advancing this field of research.
Nevertheless, CTAO operates in a pointed-observation mode with a comparatively narrow instantaneous field of view of $\approx 4$--$8^{\circ}$, which inherently lowers the probability of serendipitously catching spatially unpredictable transients such as PBH evaporation bursts.

The remainder of this paper is organized as follows.
In Section~\ref{sec:2}, we present the theoretical framework and methodology for searching for PBH bursts.
Section~\ref{sec:4} outlines the expected sensitivities of CTAO in measuring the local burst rate density.
Finally, Section~\ref{sec:5} provides the conclusions of this study.

\section{Methods for detecting PBH bursts}\label{sec:2}

\subsection{Gamma-ray photon spectrum}

Based on the concept of Hawking radiation, the relationship between the temperature $T$ of a non-rotating black hole and its mass $M$ is expressed as \cite{Hawking:1974rv}
\begin{equation} T = \frac{M_p^2}{8 \pi M},
\end{equation}
where  {$M_p$} denotes the Planck mass. As a black hole undergoes evaporation, its mass decreases, while its temperature increases. This results in an increasing variety and quantity of particles being emitted by the black hole. In the final stages of the evaporation process, a significant outburst of particles is released, resulting in a burst event. The particles emitted by Hawking radiation are considered to consist solely of particles from the Standard Model, denoted as the standard evaporation model (SEM) \cite{MacGibbon:1990zk,MacGibbon:1991tj}. In addition to the standard evaporation model (SEM) itself, a number of extensions to the SEM have been proposed to address its possible limitations, including extra-dimension scenarios, hidden-sector emission, and supersymmetric (SUSY) particle production (see the review in Ref.~\cite{ParticleDataGroup:2022pth}). In this work we restrict our analysis to the canonical SEM.

According to the SEM theory, the relationship between the temperature $T$ and the remaining lifetime $\tau$ of a black hole can be derived from the following equation \cite{Halzen:1991uw}:
\begin{equation}
T = 7.8 \times 10^3 \left(\frac{1 \mathrm{~s}}{\tau}\right)^{\frac{1}{3}} \mathrm{GeV}\ .
\end{equation}
When $T$ exceeds 10 GeV, the time-integrated photon flux from a black hole is expressed as \cite{Petkov:2008rz} \begin{equation}\label{eq:photon spetra}
\frac{d N}{d E} = 9 \times 10^{35}\left\{\begin{aligned}
& \left(\frac{1 \mathrm{GeV}}{T}\right)^{\frac{3}{2}}\left(\frac{1 \mathrm{GeV}}{E}\right)^{\frac{3}{2}} \mathrm{GeV}^{-1}  \text { for } E<T \\
& \left(\frac{1 \mathrm{GeV}}{E}\right)^3 \mathrm{GeV}^{-1} \text { for } E \geq T
\end{aligned} \right. \ .
\end{equation}
Here, $E$ represents the energy of photons, and $N$ denotes the number of photons emitted during the period from the start of observation until the complete evaporation of the black hole.

\begin{figure}[htbp]
\centering\includegraphics[width=0.8\columnwidth]{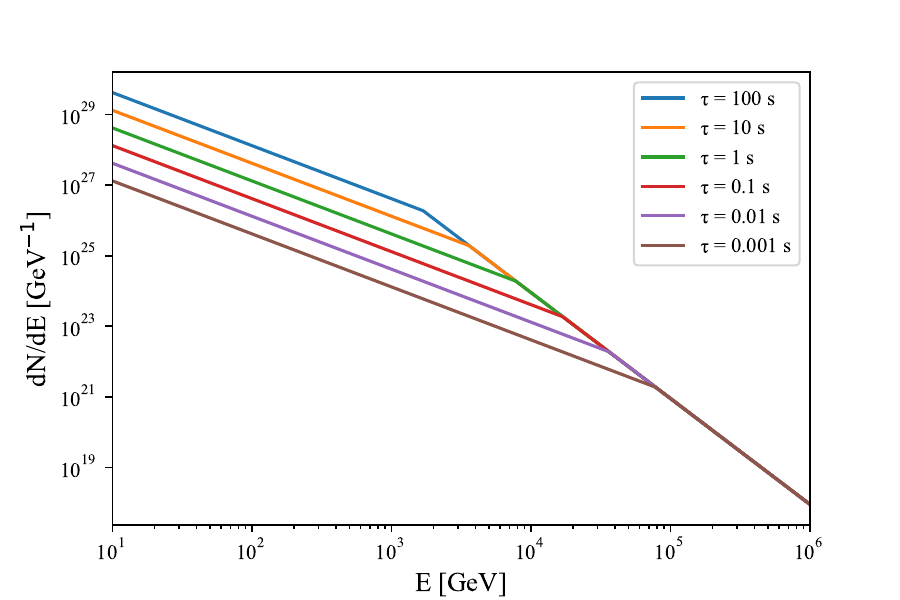}
\caption{Time-averaged gamma-ray photon spectra corresponding to the remaining lifetimes of PBHs, ranging from 0.001\,s to 100\,s.}
\label{fig:tispectra}
\end{figure}

This photon flux includes photons emitted directly by the black hole, as well as those produced by the decay and final-state radiation of other species emitted directly. When ${E} \geq {T}$, the flux is primarily governed by photons directly emitted via Hawking radiation, while for ${E} < {T}$, the flux is mainly influenced by the decay and final-state radiation of other particles. Figure~\ref{fig:tispectra} shows the time-integrated gamma-ray photon spectra for various remaining lifetimes of PBHs, ranging from 0.001\,s to 100\,s, as used in this study.

To constrain the density of local PBH bursts, it is necessary to calculate the expected number of photons, ${\mu(r, {\theta_i}, \tau)}$, detected from a PBH burst. This can be computed using the following equation: \begin{equation} \label{eq:mu}
\mu(r, \theta_i, \tau)=\frac{(1-f)}{4 \pi r^2} \int_{E_1}^{E_2} \frac{d N(\tau)}{d E} A(E, \theta_i) d E\ ,
\end{equation}
where ${A(E, \theta_i)}$ represents the detector's effective area, ${f}$ denotes the detector's dead time, $r$ is the distance, and $\theta_i$ represents the zenith angle of band $i$. Additionally, ${E_1}$ and ${E_2}$ correspond to the lower and upper boundaries of the detector's energy range, respectively.

\subsection{Method of data analysis}

To claim a detection of PBH bursts or place upper limits on the local burst-rate density, it is necessary to establish a detection criterion for data analysis.

For a null detection of PBH bursts, all photons received by the detector are attributed to the background. The number of received photons generally follows a Poisson distribution, with the expected value given by the number of background photons $n_\mathrm{bk}$. If the number of received photons exceeds a threshold $n_c$, the observed signal is no longer consistent with background fluctuations, but rather indicates a burst event. The p-value corresponding to this threshold $n_c$ is defined as $p_c$,
\begin{equation}\label{eq:possion} p_c = P\left(\geq n_c \mid n_\mathrm{bk}\right) . \end{equation}
Upon detection of a burst signal, the number of photons received by the detector follows a new Poisson distribution with an expected value of $n_\mathrm{bk}+\mu(r, \theta_i, \tau)$, where $\mu(r, \theta_i, \tau)$ denotes the expected number of photons from the burst.

In an externally triggered search, where the time and location of the event are known, the threshold p-value $p_c$ is typically set to correspond to a $5\sigma$ significance level. However, in a blind search, the time and location of the burst are unknown, requiring a search over all possible time and spatial windows. This process introduces a large number of trials, which must be accounted for when defining the detection criterion. To correct for the trials, the threshold p-value $p_c$ is adjusted according to the number of trials $N_t$, following Ref.~\cite{Ukwatta:2015iba}.

The p-value corresponding to a $5\sigma$ significance, typically set as $p_{0}=2.3\times10^{-7}$, is related to the single-trial p-value $p_c$ through the relation
\begin{equation}\label{eq:correction} 
p_0 = 1 - (1 - p_c)^{N_{\mathrm{t}}} .
\end{equation}

Here, \(p_c\) denotes the probability of a false alarm in a single, independent time–space window, while \(p_0\) is the global false-alarm probability across the entire search. Assuming statistical independence among all \(N_t\) trials, the probability that none of them exceeds the detection threshold is \((1 - p_c)^{N_t}\). The complement gives the probability of at least one false alarm occurring, yielding Eq.~(\ref{eq:correction}). In the limit of small \(p_c\), a first-order expansion leads to the commonly used approximation
\begin{equation}
p_c \approx \frac{p_0}{N_t} ,
\end{equation}
which is valid when \(N_t p_c \ll 1\) and widely adopted in high-significance searches.

The total number of independent trials \(N_t\) is estimated following Ref.~\cite{Lopez-Coto:2022tcg}:
\begin{equation} 
N_{\mathrm{t}}(\tau) = \frac{S}{\tau} \left( \frac{\theta_{\text{fov}}}{\theta_{\text{res}}} \right)^2 ,
\end{equation}
where \(S\) is the total search duration, \(\theta_{\text{fov}}\) the field of view of the detector, and \(\theta_{\text{res}}\) its angular resolution. 

To calculate the upper limit on the local PBH burst rate density, the minimum number of photons $\mu_{\min}(r, \theta_i, \tau)$ expected from a PBH burst must first be estimated. In a standard search \cite{Ukwatta:2015iba, HAWC:2019wla}, it is assumed that when the number of detected photons exceeds or equals the criterion $n_c$, the p-value of a new Poisson distribution with expectation of $n_\mathrm{bk}+\mu_{\min}(r, \theta_i, \tau)$ is 0.5. The value of $\mu_{\min}(r, \theta_i, \tau)$ can be obtained by solving the equation
\begin{equation} \label{eq:P} P(\geq n_c \mid n_\mathrm{bk}+\mu_{\min}(r,\theta_i, \tau))=0.5\ .
\end{equation}
By substituting $\mu_{\min}(r, \theta_i, \tau)$ into Eq.~(\ref{eq:mu}), the maximum detectable distance of a PBH burst by the detector can be determined as
\begin{equation}\label{eq:radius} r_{\max }\left(\theta_i, \tau\right)=\sqrt{\frac{(1-f)}{4 \pi \mu_{\min}\left(\theta_i, \tau\right)} \int_{E_1}^{E_2} \frac{d N(\tau)}{d E} A(E, \theta_i) d E}\ ,
\end{equation}
where $\theta_i$ denotes the zenith angle band labeled by $i$. The total detectable volume of PBH for the detector is given by
\begin{equation}\label{eq:volume} V(\tau)=\sum_i V\left(\theta_i, \tau\right)=\frac{4}{3} \pi \sum_i r_{\max }^3\left(\theta_i, \tau\right) \frac{\mathrm{FOV}\left(\theta_i\right)}{4 \pi}\ ,
\end{equation}
with the solid angle corresponding to the field of view expressed as
\begin{equation}\label{eq:fov} \mathrm{FOV}\left(\theta_i\right)=2 \pi\left(\cos \theta_{i, \min }-\cos \theta_{i, \max }\right)\ ,
\end{equation}
where $\theta_{i, \min }$ and $\theta_{i, \max }$ represent the minimum and maximum zenith angles of band $i$, respectively.

In the absence of detected PBH bursts, a uniform spatial distribution of PBHs in the solar neighborhood is assumed. The Poisson probability of observing zero PBH bursts at a confidence level $X$ is given by
\begin{equation}
P(0 \mid m)=1-P(n \geq 1 \mid m)=1-X=\frac{m^0 e^{-m}}{0 !}\ ,
\end{equation}
where $m$ represents the expected number of PBH bursts. By setting $X = 99\%$, the corresponding $m$ is obtained as $\ln 100 \approx 4.6$. Consequently, the upper limit of the PBH burst rate density is given by~\cite{Abdo:2014apa}: \begin{equation} U L_{99}=\frac{4.6}{V S}\ , \end{equation} where $S$ denotes the total search duration and $V$ is the total detectable volume, as evaluated in Eq.~(\ref{eq:volume}).

\begin{table}[!h]
\centering
\caption{\label{tab:table2}The maximum detectable distance and PBH burst local rate density upper limit in 3 and 5 years for various remaining PBH lifetimes for CTAO south array and CTAO north array.}
\begingroup
\setlength{\tabcolsep}{4.2pt}        
\renewcommand{\arraystretch}{1.12}    
\begin{tabular}{@{}l c c c c c@{}}  
\toprule
Detector & $\tau$ (s) & 3-yr $r_\mathrm{max}$ (pc) & 5-yr $r_\mathrm{max}$ (pc) & 3-yr limit ($\mathrm{pc}^{-3}\mathrm{yr}^{-1}$) & 5-yr limit ($\mathrm{pc}^{-3}\mathrm{yr}^{-1}$) \\
\midrule
\multirow{6}{*}{CTAO (South)}
 & $10^{-3}$ & 0.312 & 0.312 & 11258.54 & 6755.12 \\
 & $10^{-2}$ & 0.540 & 0.540 & 2171.52  & 1302.91 \\
 & $10^{-1}$ & 0.821 & 0.821 & 617.90   & 370.74  \\
 & 1         & 1.157 & 1.157 & 220.77   & 132.46  \\
 & 10        & 1.507 & 1.507 & 99.91    & 59.95   \\
 & $10^{2}$  & 1.782 & 1.782 & 60.43    & 36.26   \\
\midrule
\multirow{6}{*}{CTAO (North)}
 & $10^{-3}$ & 0.218 & 0.218 & 100382.45 & 60229.42 \\
 & $10^{-2}$ & 0.341 & 0.341 & 26227.94  & 15736.76 \\
 & $10^{-1}$ & 0.541 & 0.541 & 6568.04   & 3940.82  \\
 & 1         & 0.802 & 0.802 & 2016.06   & 1209.64  \\
 & 10        & 1.070 & 1.070 & 848.94    & 509.36   \\
 & $10^{2}$  & 1.250 & 1.250 & 532.47    & 319.48   \\
\bottomrule
\end{tabular}
\endgroup
\end{table}
\section{Sensitivities of CTAO in measurements of local burst-rate density}\label{sec:3}

Detecting PBH bursts in the local universe presents a formidable challenge, requiring advanced gamma-ray detectors. CTAO, as one of the most advanced gamma-ray observatories at present and in the foreseeable future, is well-equipped to undertake this mission. In this section, we present the upper limit results on the PBH burst rate density achievable by CTAO in the local universe.

As the next-generation ground-based gamma-ray observatory, CTAO employs the imaging atmospheric Cherenkov telescope (IACT) technique and primarily consists of three types of telescope: small-sized telescope (SST) observing in the energy range of 5 TeV to 300 TeV, medium-sized telescope (MST)  observing in the energy range of 150 GeV to 5 TeV, and large-sized telescope (LST) observing in the energy range of 20 GeV to 150 GeV. CTAO plans to construct two arrays. The northern array will be located on the island of La Palma in Spain and will consist of 4 SSTs and 9 MSTs, mainly for observing extragalactic gamma-ray signals. The southern array will be located in the Atacama Desert in Chile and will consist of 16 MSTs and 17 SSTs, mainly to observe galactic gamma-ray signals \cite{CTAConsortium:2017dvg}.

Since the CTAO consists of two telescope arrays located in the northern and southern hemispheres, each array is considered a separate detector to evaluate the constraints on the local PBH burst rate density. Compared to most detectors, which are only sensitive to high-energy gamma rays, the CTAO is sensitive to both high-energy and low-energy gamma rays, making it more effective in constraining local PBH bursts with long remaining lifetimes

To calculate the sensitivity of CTAO to local PBH bursts, we used the method introduced in Section \ref{sec:2}. First, we calculate the number of background particles ${n_\mathrm{b k}}$. The background particle rate ${R_b\left(\theta_i\right)}$ is provided in Ref.~\cite{CTA}. Next, we calculate the maximum detection distances of CTAO, denoted as ${ r_{\max }\left(\theta_i, \tau\right)}$, using Eqs.~(\ref{eq:P}) and (\ref{eq:radius}). The photon effective area, background particle rate, and field of view of CTAO used in these calculations are also obtained from Refs.~\cite{CTA,CTA1}. In this analysis, we adopted the 50-hour photon effective area of CTAO after gamma/hadron separation.  We also assume that the dead time can be neglected in this calculation.

Finally, we determine the effective detection volume using ${r_{\max }\left(\theta_i, \tau\right)}$ and use the volume to estimate the upper limits of the local PBH burst density for the CTAO south and north arrays at 99\% confidence level. The results of the maximum detection distance of PBH and the upper limit density of the local
PBH burst rate in 3 and 5 years for CTAO south array and CTAO north array are listed in Table~\ref{tab:table2}.

It should be noted that our calculations are based on an idealised on-axis instrumental response. According to the public CTAO point-source performance curves, off-axis effects can reduce the sensitivity by approximately a factor of two \cite{ctao_irf_prod5_v0.1}. As this does not affect our main conclusions, we restrict our analysis to this ideal case.

In this section, we compare our results from Table~\ref{tab:table2} with those of other studies and discuss several issues encountered during the analysis process.

\begin{figure}[!h]
\centering\includegraphics[width=0.8\columnwidth]{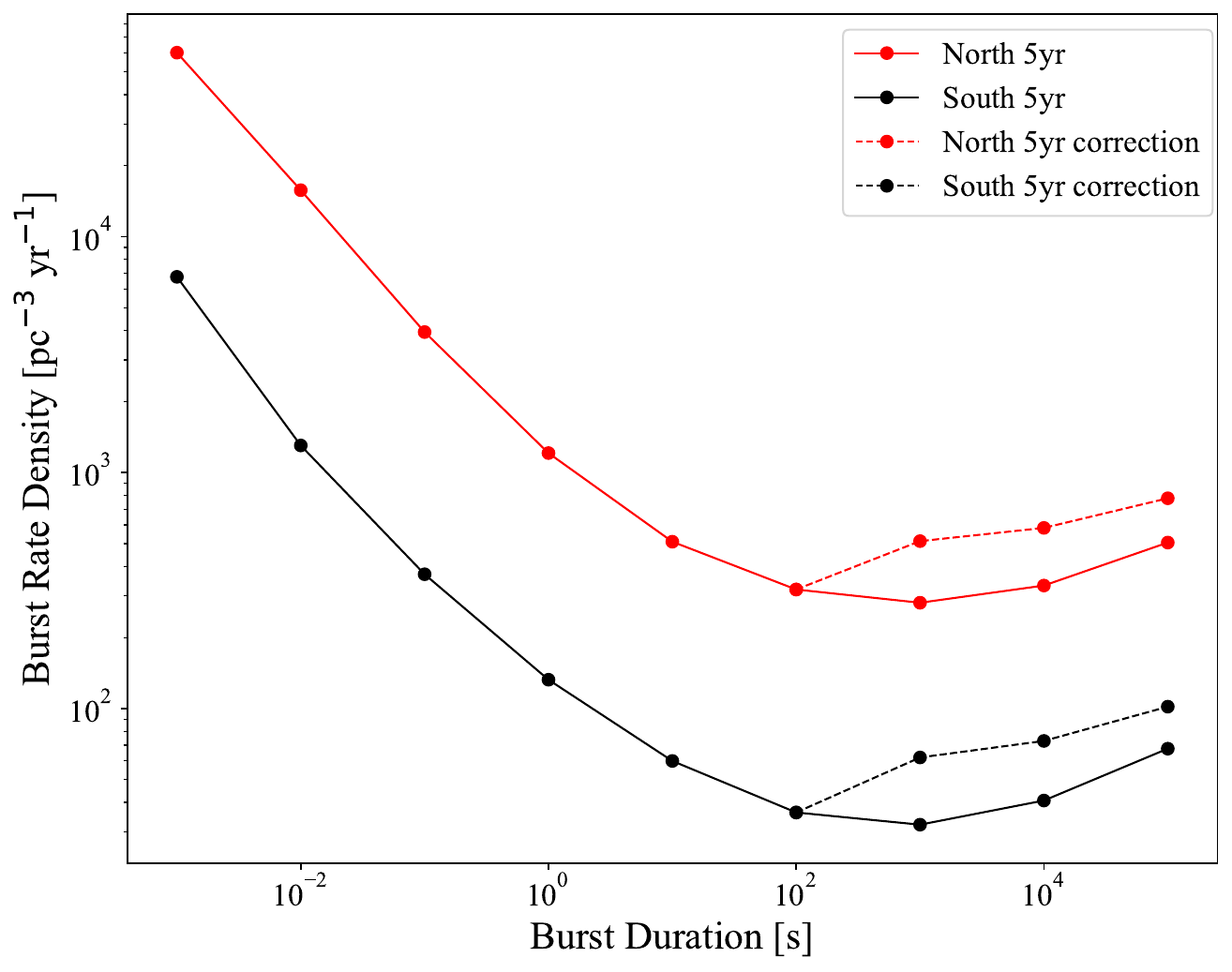}
\caption{5-year limit results of CTAO with and without correction. }
\label{fig:fig10}
\end{figure}
\section{Discussion}\label{sec:4}

\begin{figure*}[h]
\centering\includegraphics[width=1\textwidth]{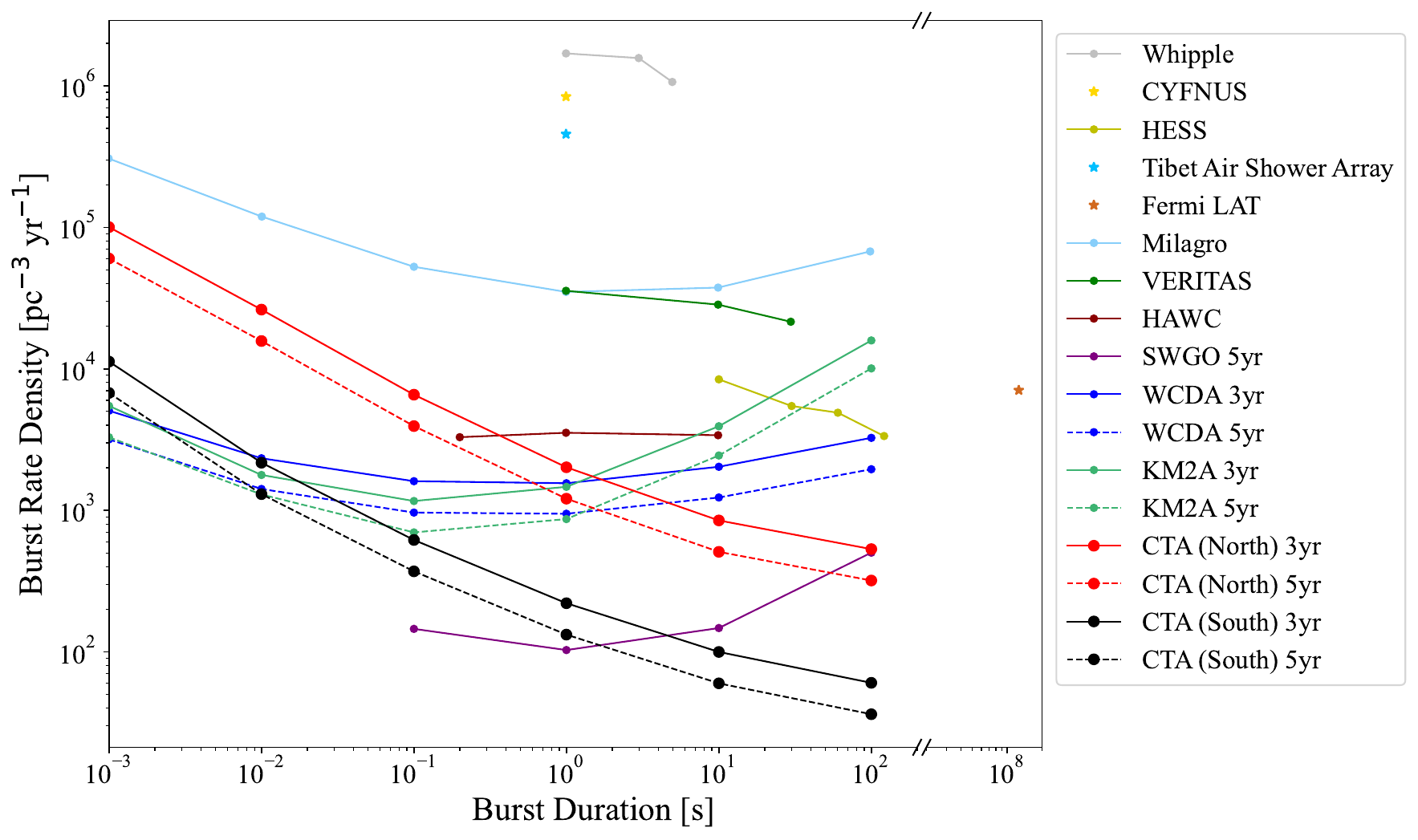}
\caption{The PBH burst local rate density upper limits of CTAO at 99\% confidence level, compared with the results from Whipple \cite{Linton:2006yu}, CYGNUS \cite{Alexandreas:1993zx}, H.E.S.S. \cite{HESS:2023zzd}, MAGIC \cite{Cassanyes:2015wpr}, CTAO2015 \cite{Mukherjee_2018},Tibet Air Shower Array \cite{amenomori1995search}, Milagro \cite{Abdo:2014apa}, VERITAS \cite{Archambault:2017asc}, Fermi-LAT \cite{Fermi-LAT:2018pfs}, HAWC \cite{HAWC:2019wla}, LHAASO \cite{Yang:2024vij} and SWGO \cite{Lopez-Coto:2022tcg}.}
\label{fig:figure8}
\end{figure*}

Figure~\ref{fig:figure8} presents a comparison between our constraint results and those obtained by other detectors based on observational data \cite{Linton:2006yu,Alexandreas:1993zx,amenomori1995search,Archambault:2017asc,Abdo:2014apa,Fermi-LAT:2018pfs,HAWC:2019wla,HESS:2023zzd,Lopez-Coto:2022tcg,Yang:2024vij}. It is evident that CTAO exhibits significantly higher sensitivity in constraining the local PBH burst rate density compared to most previous detectors. The sensitivity of CTAO is only slightly lower than that of LHAASO for bursts with durations shorter than 0.01\,s and lower than that of the future SWGO detector for bursts with durations shorter than 1\,s. This demonstrates that CTAO holds a distinct advantage in searching for long-duration PBH bursts.

To establish a reference for selecting the optimal burst duration when searching for PBH bursts in future observational studies, we analyze the burst durations to which CTAO is most sensitive. This analysis is a key focus of our work. As shown in Table~\ref{tab:table2}, better constraints are likely achieved by targeting PBHs with longer remaining lifetimes. However, since telescopes typically do not observe the same region for extended periods, PBHs with remaining lifetimes significantly longer than the observation time emit more photons than can be detected. Considering this limitation, we apply a rough correction for observations lasting longer than $10^2$\,s to determine the optimal constraints on PBH bursts for CTAO. Here, the value of $10^2$ s refers to the integration window $\Delta t$ used in the PBH burst-search algorithm, rather than the total duration of an observational run. A typical run lasts about $1200$ s for CTA or approximately $28$ minutes for H.E.S.S.~\cite{HESS:2023zzd}. Following previous analyses, such as Ref.~\cite{HESS:2021rto}, we adopt $\Delta t = 10^2$ s to achieve a balance between signal retention and background accumulation.

We extended the analysis to include remaining lifetimes up to $10^5$\,s. This extension serves two purposes by identifying the burst duration at which CTAO is most sensitive and evaluating the observational limitations related to PBHs with long remaining lifetimes.

For PBHs with remaining lifetimes of $10^3$\,s and $10^4$\,s, we assume that only 50\% of the total emitted photons are detected, as the burst duration exceeds the typical integration time. For $\tau = 10^5$\,s, we apply an additional correction due to the limited nighttime visibility.
Here, we define time in terms of the PBH remaining lifetime $\tau$, i.e., the time until complete evaporation. Since telescopes cannot observe continuously over such long timescales, we conservatively assume that the observation spans two nights. To maximize realism, we adopt the shortest night of the year as the available window.
For the first night, we assume the observation begins when the PBH has $\tau = 10^5$\,s remaining and ends at $\tau = 8.94 \times 10^4$\,s, based on the shortest possible dark time. For the second night, the observation window extends from $\tau = 2.93 \times 10^4$\,s to $\tau = 0$\,s for the South array, and from $\tau = 2.85 \times 10^4$\,s to $\tau = 0$\,s for the North array. This corresponds to the maximum usable duration of the shortest night at each site.
Combining these two nights covers approximately 11 hours of the 28-hour-long PBH burst. We therefore apply an additional 50\% correction to the photon counts for $\tau = 10^5$\,s. The resulting 5-year burst rate density limits for CTAO, before and after applying this correction, are shown in Fig.~\ref{fig:fig10}.

From Fig.~\ref{fig:fig10}, we observe that, without applying the correction, CTAO imposes the strongest constraints on PBH burst rate density with a remaining lifetime of $10^3$\,s. Therefore, to achieve the most stringent limits, CTAO should consider setting the observation duration for the same sky region to exceed $10^3$\,s.

\section{Conclusions}\label{sec:5}

PBHs, formed in the early universe, are significant candidates for dark matter. Searching for PBHs can enhance our understanding of both the origin and evolution of the universe, as well as the nature of dark matter. This provides the primary motivation for searching local PBH bursts with CTAO.

In this paper, we first estimate the number of photons detected by CTAO from local PBH bursts, based on detection criteria and background particle rates. We then used the number of detected photons and CTAO's effective area to calculate the maximum detection distance for local PBH bursts. Finally, we estimated the upper limit of the burst rate density, in the situation that no local PBH bursts are detected.

In summary, we showed that CTAO can significantly enhance the detection capabilities of PBHs on the parsec scale. We found that the best constraint on the local PBH burst rate density for CTAO is $\dot{\rho}< 36.26 ~\mathrm{pc}^{-3} \mathrm{yr}^{-1}$. This result is one to two orders of magnitude better than the constraint obtained by LHAASO, as predicted in Ref.~\cite{Yang:2024vij}, and the current strongest constraints obtained by HAWC. Even when compared to SWGO, another future detector, the best constraint results are several times more stringent. Furthermore, we found that when the observation time for the same sky region is extended to $10^3$\,s, CTAO can achieves the best constraint results. These findings can help optimize detection strategies for future observations to obtain better results. We hope that, in the future, by combining actual observational data from CTAO and optimizing observation methods such as considering different search bin sizes, gamma-hadron separation parameters, and selection cuts for different burst durations, as discussed in Ref.~\cite{Abdo:2014apa}, we can obtain even better constraints.

\begin{acknowledgments}
This work was supported by the National SKA Program of China (Grants Nos. 2022SKA0110200 and 2022SKA0110203), the National Natural Science Foundation of China (Grants Nos. 12533001, 12473001, 11975072, 11875102, and 11835009), the China Manned Space Program (Grant No. CMS-CSST-2025-A02), and the 111 Project (Grant No. B16009).
\end{acknowledgments}

\bibliographystyle{JHEP}
\bibliography{paper}
 
 \end{document}